\documentclass[pdflatex,sn-basic,iicol]{sn-jnl}

\usepackage{graphicx}%
\usepackage{multirow}%
\usepackage{amsmath,amssymb,amsfonts}%
\usepackage{amsthm}%
\usepackage{mathrsfs}%
\usepackage[title]{appendix}%
\usepackage{xcolor}%
\usepackage{textcomp}%
\usepackage{manyfoot}%
\usepackage{booktabs}%
\usepackage{algorithm}%
\usepackage{algorithmicx}%
\usepackage{algpseudocode}%
\usepackage{listings}%

\usepackage{tabularx}%

\usepackage{pgfplots}
\pgfplotsset{compat=newest}
\usetikzlibrary{plotmarks}
\usetikzlibrary{arrows.meta}
\usepgfplotslibrary{patchplots}
\usepackage{grffile}
\usepackage{amsmath}
\newlength\fwidth
\newlength\fheight
\newlength\fsep
\newlength\labsep

\usepgfplotslibrary{external} 
\usetikzlibrary{calc}

\begin{document}


\title{Experimental assessment of square wave spatial spanwise forcing of a turbulent boundary layer}


\author*[1]{\fnm{Max W.} \sur{Knoop}}\email{m.w.knoop@tudelft.nl}

\author[1,2]{\fnm{Friso H.} \sur{Hartog}}

\author[1]{\fnm{Ferdinand F. J.} \sur{Schrijer}}
\author[1,2]{\fnm{Olaf W. G.} \sur{van Campenhout}}
\author[1,2]{\fnm{Michiel} \sur{van Nesselrooij}}
\author[1]{\fnm{Bas W.} \sur{van Oudheusden}}

\affil[1]{\orgdiv{Faculty of Aerospace Engineering}, \orgname{Delft University of Technology}, \orgaddress{\street{Kluyverweg 1},  \postcode{2629 HS}, \city{Delft} ,\country{The Netherlands}}}

\affil[2]{\orgname{Dimple Aerospace B.V}, \orgaddress{\street{Kluyverweg 1},  \postcode{2629 HS}, \city{Delft} ,\country{The Netherlands}}}

\abstract{
We present an experimental realisation of spatial spanwise forcing in a turbulent boundary layer flow, aimed at reducing the frictional drag. The forcing is achieved by a series of spanwise running belts, running in alternating spanwise direction, thereby generating a steady spatial square-wave forcing. Stereoscopic particle image velocimetry in the streamwise-wall-normal plane is used to investigate the impact of actuation on the flow in terms of turbulence statistics, drag performance characteristics, and spanwise velocity profiles, for a non-dimensional wavelength of $\lambda_x^+ = 397$. 
In line with reported numerical studies, we confirm that a significant flow control effect can be realised with this type of forcing. The scalar fields of the higher-order turbulence statistics show a strong attenuation of stresses and production of turbulence kinetic energy over the first belt already, followed by a more gradual decrease to a steady-state energy response over the second belt. The streamwise velocity in the near-wall region is reduced, indicative of a drag-reduced flow state. The profiles of the higher-order turbulence statistics are attenuated up to a wall-normal height of $y^+ \approx 100$, with a maximum streamwise stress reduction of 45\% and a reduction of integral turbulence kinetic energy production of 39\%, for a non-dimensional actuation amplitude of $A^+ = 12.7$.
An extension of the classical laminar Stokes layer theory is introduced, based on the linear superposition of Fourier modes, to describe the non-sinusoidal boundary condition that corresponds to the current case. The experimentally obtained spanwise velocity profiles show good agreement with this extended theoretical model. The drag reduction was estimated from a linear fit in the viscous sublayer in the range $2 \leq y^+\leq 5$. The results are found to be in good qualitative agreement with the numerical implementations of \citet[Phys. Fluids, vol. 21,][]{viotti_streamwise_2009}, matching the drag reduction trend with $A^+$, and reaching a maximum of 20\%. }

\maketitle

\section{Introduction}
Spanwise forcing of the turbulent boundary layer through lateral wall motion is an active flow control technique known especially for its potential to reduce skin friction. The method involves imposing a spatio-temporal, in-plane wall oscillation in the spanwise direction. Both temporal and spatio-temporal forcing, referred to as oscillating wall (OW) and travelling wave (TW) forcing, respectively, have been extensively researched, in both numerical and experimental investigations. Drag reduction (DR) values of over 40\% and positive theoretical net power savings (NPS) have been reported at moderate friction Reynolds numbers of $Re_\tau = 200-2000$ \citep{ricco_review_2021}. DR is defined as the percentage difference in wall-shear stress ($\tau_w$) with respect to the non-actuated case. NPS is the difference between the power saving to drive the flow in the streamwise direction, and the theoretical input power to drive the spanwise forcing, expressed as a percentage with respect to the non-actuated case. The characteristic parameters are commonly expressed in viscous units; with the kinematic viscosity $\nu$ and the friction velocity $u_\tau \equiv \sqrt{\tau_w/\rho}$, with $\rho$ as the fluid density, used as scaling parameters. Furthermore, we distinguish two scaling approaches, using either the reference $u_{\tau0}$ from the non-actuated case or the actual $u_{\tau}$ of the drag-reduced flow case; these are denoted by the superscripts `+' and `*', respectively. The friction Reynolds number is defined as $Re_\tau \equiv \delta u_{\tau0}/\nu$, where $\delta$ is the boundary layer thickness. In this study, the $x$, $y$, and $z$ coordinates represent the streamwise, wall-normal, and spanwise directions, respectively, corresponding to instantaneous velocity components $u$, $v$, and $w$. An overbar and prime denote the respective mean and fluctuating components.

Early work on spanwise forcing focused on the oscillating wall (OW) actuation scheme, where the entire wall section is subjected to a temporal oscillation. The strategy was first studied using direct numerical simulations (DNS) by \citet{jung_suppression_1992}. Their simulations were conducted at a Reynolds number of $Re_\tau = 200$; the standard Reynolds number utilised in the majority of numerical investigations. Unless otherwise specified, $Re_\tau = 200$ is the considered Reynolds number for all discussed numerical work. A maximum DR of 40\% was found at a spanwise velocity amplitude of $A^+ = 12$ ($A^+ \equiv A/u_{\tau0}$) and an optimum period of $T^+ = 100$ ($T^+ \equiv T u_{\tau0}/\nu$). \citet{laadhari_turbulence_1994} conducted experiments which corroborated the results of \citet{jung_suppression_1992}. They reported a DR of 36\% using the same actuation parameters. The parametric DNS study by \citet{quadrio_critical_2004} found that DR monotonically increases at a decreasing rate with increasing amplitude, while revealing the optimum period to be $T^+ = 100-125$. 

Traveling wave (TW) forcing was introduced by \citet{quadrio_streamwise-travelling_2009}; their DNS study revealed the complex dynamics of the DR in wavenumber-frequency space at $A^+ = 12$. The results highlighted superior performance in upstream-travelling waves, with a maximum DR of 45\% and a notable NPS of 18\%. Inspired by their work, several experimental realisations of TW forcing have been made. The work of \citet{auteri_experimental_2010} realised TW forcing in their pipe flow setup, where the wave was generated by means of multiple azimuthally rotating pipe segments. Other implementations are the deformable Kagome lattice of \citet{bird_experimental_2018}, and a more recent contribution of \citet{marusic_energy-efficient_2021}, who employed a series of synchronised spanwise oscillatory wall elements. 

As a result of the wall motion, an oscillatory spanwise velocity profile emerges, which predominantly resides within the viscous region of the boundary layer. The profiles show a close match to the laminar solution of the second Stokes problem, establishing a temporal Stokes layer (TSL) \citep{stokes_effect_1851} in the case of OW forcing. In spatio-temporal form, the solution is described by the generalised Stokes layer (GSL), as introduced by \citet{quadrio_laminar_2011}. The mechanism responsible for the DR is not precisely known, but one of the paradigms suggests a favourable interaction between the Stokes layer dynamics and near-wall coherent structures. \citet{ricco_modification_2004} and \citet{kempaiah_3-dimensional_2020} present evidence that the Stokes layer disrupts this interaction by laterally displacing near-wall streaks and quasi-streamwise vortices, disrupting their spatial coherence and weakening the drag-producing near-wall cycle. Moreover, the work of \citet{Bradshaw_measuements_1985} and \citet{Agostini_spanwise_2014} support evidence that high rates of change of Stokes strain, not the Stokes stain itself ($\partial\overline{w}/\partial y$) drives the drag reduction, corresponding to the phase where the wall motion changes direction.

A third spanwise actuation scheme is a pure spatial forcing, which has been the focus in a small number of numerical studies \citep[e.g][]{viotti_streamwise_2009, yakeno_spatio-temporally_2009, skote_comparison_2013}. Forcing is imposed in this case in the form of a standing wave (SW), by applying a steady wall velocity that varies in the streamwise direction: 
\begin{equation}
\label{eq:SW}
    W_w(x) = A \sin \left( \frac{2\pi}{\lambda_x}x\right),
\end{equation}
in which $A$ is the spanwise velocity amplitude and $\lambda_x$ the streamwise wavelength of actuation. The oscillatory spanwise forcing of the flow is in this concept achieved convectively, as the flow travels over the regions of alternating forcing direction. SW forcing was first introduced and numerically studied by \citet{viotti_streamwise_2009}, revealing an optimum DR of 45\% around $\lambda_x^+ = 1,000-1,250$ ($\lambda_x^+ \equiv \lambda_x u_{\tau0}/\nu$) at $A^+ = 12$. The optimum NPS of 23\% is found at a lower amplitude of $A^+ = 6$, resulting from the balance between DR and input power requirements. Although SW forcing was not their primary focus, the experimental study of \citet{auteri_experimental_2010} realised spatial forcing when their pipe flow setup was actuated under steady conditions. Apart from this study, no experimental realisations of SW forcing have been reported.

Despite the theoretical promise of achieving positive NPS through spanwise forcing, as an active flow control method this technique faces challenges due to its complex system architecture and substantial physical power requirements. Nevertheless, we contend that spanwise forcing could offer substantial energy-saving potential in practical applications when implemented in a passive manner, e.g. by means of oblique wavy walls \citep{ghebali_large-scale_2017} or dimpled surfaces \citep{Nesselrooij_drag_2016}. In this regard, a fundamental understanding of the underlying flow physics responsible for the drag reduction, and the important forcing mechanisms, can advance the development of passive flow control techniques for said purpose. Specifically, spatial forcing is most appropriate for this, since it is most analogous to a passive counterpart.

This work is motivated by the need to study spatial spanwise forcing at a fundamental level. Furthermore, we address the sparse attention devoted to SW forcing so far, especially experimentally. We present a proof of concept for an experimental realisation of such forcing in an external boundary layer flow. The concept was inspired by the experiment of \citet{kiesow_near-wall_2003}, who employed a single spanwise running belt to study the response of the turbulent boundary layer to a steady spanwise crossflow. Extending their idea, the turbulent boundary layer is forced by a series of belts running in alternating spanwise directions, so as to generate a steady spatial square-wave wall boundary condition. 
The objective of the current study is to present the experimental concept and to subsequently discuss its associated flow control potential. The flow modulation effects are characterised using stereoscopic particle image velocimetry (SPIV), in terms of the turbulence statistics, an extended model of the classical SSL theory to account for the non-sinusoidal waveform, and its DR and NPS performance.
\section{Experimental methodology}
\begin{figure*}
    \centering
    \includegraphics[]{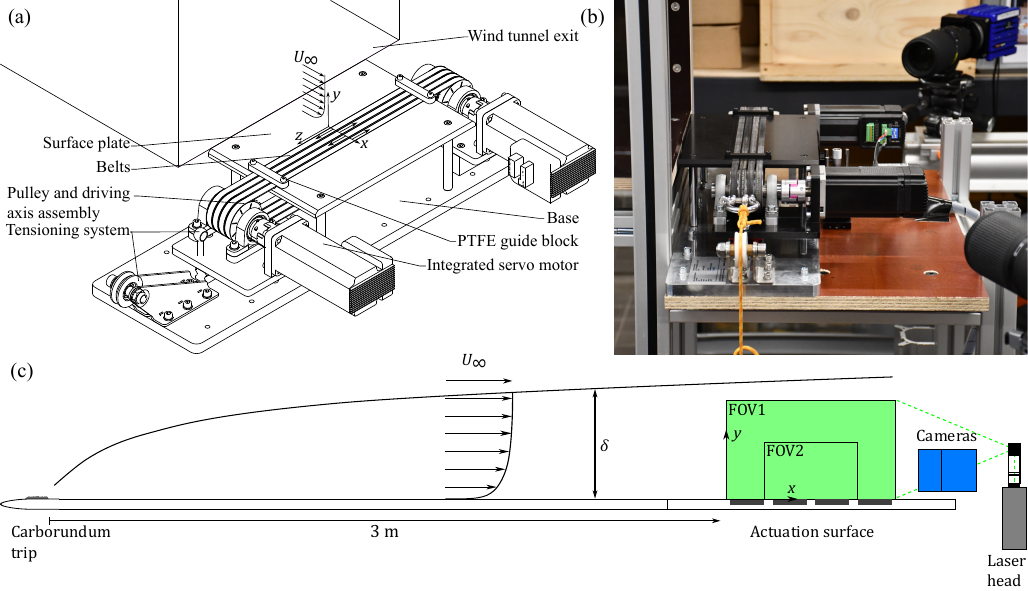}
    \caption{(a) Schematic representation of the SSES mechanism, and its placement with respect to the wind tunnel exit. Arrows on the belts indicate their respective rotation directions. (b) Photograph of the experimental configuration. (c) Graphical representation of the SPIV experiment.}
    \label{fig:prototype_schematic}
\end{figure*}

Experiments were performed in a low-speed wind tunnel of the Delft University of Technology at a freestream velocity of $U_\infty \approx 6.7$ m/s and a freestream turbulence intensity of the order of 0.7\%. The open-return tunnel has a test section with a cross-section of 0.4 m \texttimes\ 0.4 m. A clean turbulent boundary layer was generated on a horizontally oriented flat plate with an elliptical leading edge and tripped using carborundum roughness (24 grit), achieving a turbulent development length of 3 m. The experimental setup was placed in an open-jet configuration, behind and flush with the test section. A graphical representation and photograph of the experimental configuration is given in Fig.~\ref{fig:prototype_schematic}. A temperature sensor and Pitot-static tube were used to monitor ambient conditions and freestream velocity. Given the limited field of view (FOV) of the SPIV experiment, the reference boundary layer thickness under non-actuated conditions was characterised over its full height using hot-wire anemometry. The hot-wire probe was positioned 3 mm downstream of the actuation surface. The boundary layer characteristics for the current experiment are $\delta = 67$ mm and $u_{\tau0} = 0.27$ m/s, yielding a friction Reynolds number of $Re_\tau = 1230$.

\subsection{Steady spanwise excitation setup}
The experimental setup, that will be referred to as the steady spanwise excitation setup (SSES), was designed to realise SW forcing experimentally in an external boundary layer flow. Four streamwise-spaced, spanwise running belts, run in alternating direction so as to generate a spatial square waveform. This waveform, along with other non-sinusoidal waveforms, has been previously investigated for OW forcing using DNS by \citet{cimarelli_prediction_2013}. Their study revealed that the square wave outperformed all other waveforms in terms of DR. However, in terms of NPS, the sinusoidal waveform still showed optimal performance, as a result of the substantially higher power requirements for the square-wave forcing. The use of a half square waveform was investigated numerically by \citet{mishra_drag_2015} motivated by their potential for reduced power requirements. From that point of view, the optimal NPS of 18\% was found at $A^+=10$, which is marginally lower than the optimum 23\% found by \citet{viotti_streamwise_2009}.

A schematic representation of the experimental apparatus and its constituting components are presented in Fig.~\ref{fig:prototype_schematic}. The experimental setup is equipped with four 9 mm wide neoprene belts, extending over a spanwise extent of 256 mm. The streamwise separation of 2 mm between the belts is kept to a minimum while considering the mechanical feasibility of the design. We define the start of the waveform, and the origin of the coordinate system, at 1 mm (i.e. half the streamwise separation) upstream of the first belt. The belts are embedded in an aluminium surface plate so as to run flush with the wall. In addition, the belts are guided by a PTFE strip to constrain the belts to the surface grooves.  The grooves are precisely machined to strict tolerances, with a maximum gap and step size of 50 \textmu m ($0.9\nu/u_\tau$).
The belts are driven by a non-slip pulley system, which allows for individually controlling the rotation direction of the belts. An advantage of the current experimental setup is the independent control over the actuation parameters, $A^+$ and $\lambda_x^+$, whereas the amplitude and period are coupled in OW and TW forcing that is realised using oscillatory elements at a fixed displacement amplitude. $A^+$ is directly controlled by setting the belt speed, while $\lambda_x^+$ can be changed by either altering the viscous scaling parameters (notably, the friction velocity through the freestream speed) or through changing the physical wavelength by adapting the rotation directions of the individual belts. It is worth noting that the latter option introduces a non-perfect square waveform due to the non-actuated transition regions of 2 mm between belts. Due to the limited extent of the actuation surface, a parametric study into changing the physical wavelength is considered beyond the scope of the current research. The experimental apparatus is powered by two iHSV60 integrated AC servo motors that drive the belts in either positive or negative spanwise direction. The motors have an adjustable feedback control loop for a fixed belt velocity output, with a maximum spanwise velocity of 6 m/s. 

The wall-normal belt vibration and displacement were characterised using a PSV-500 Poltec scanning vibrometer. The wall-normal velocity was measured at fourteen sampling points over the four belts, for 10 s at a sampling frequency of 1 kHz. The displacement was subsequently obtained by integrating the velocity signal. For the considered operational range of the experimental setup the standard deviation of the displacement was between 0.03-0.08 mm or ($0.5-1.4 \nu/u_\tau$). In addition, no significant impact of the wall-normal displacements was observed in the SPIV experiment.

\subsection{Stereoscopic particle image velocimetry}
SPIV was used to obtain the three velocity components in the streamwise-wall-normal ($x-y$) plane. The measurement details are outlined in table~\ref{tab:params}. Each measurement consists of 1000 uncorrelated velocity fields, separated by approximately 10 boundary layer turnovers ($\delta/U_\infty$), acquired at a frequency of 10 Hz. Two digital LaVision sCMOS cameras were placed on either side of the laser sheet in a back-scatter configuration. Two fields of view are considered. FOV1 was employed to assess the spatial development of various turbulence statistics, covering the four belts, spanning 46.5 mm \texttimes\ 17.5 mm, in the streamwise and wall-normal direction, respectively. To obtain ensemble averaged turbulence statistics at high spatial resolution, a smaller FOV2 was used, centred over the central two belts, with a size of 24 mm \texttimes\ 14.5 mm. Imaging was performed using Nikkor AF-S 200 mm lenses, with Kenko Teleplus PRO 300 $2\times$ teleconverters, at an aperture of f/11. The cameras were oriented at $\pm 30^\circ$ with respect to the laser sheet normal. Scheimpflug adapters were used to ensure uniform focus across the FOV. Illumination was provided by a double-pulsed Quantel Evergreen 200 laser, with a 100 mJ/pulse power setting. A combination of cylindrical and spherical lenses was used to create a laser sheet with an approximate thickness of 1 mm. Seeding was injected at the wind tunnel inlet by a SAFEX Fog 2010+ fog generator to create 1 \textmu m tracer particles from a water-glycol mixture. A graphical representation of the PIV experiment is depicted in Fig.~\ref{fig:prototype_schematic}.

Stereo calibration was performed to transform the images to physical coordinates. Additionally, self-calibration was performed to reduce the calibration fit error to below 0.25 pixel. The image pre-processing involved two procedures, subtracting the minimum over time from a set of 5 images, and a normalisation with the local average using a 10-pixel kernel. For each camera, the instantaneous two velocity component fields were obtained using cross-correlation, with a 16 \texttimes\ 16 pixels\textsuperscript{2} Gaussian windowing function at 75\% overlap. To further increase the near-wall resolution for the determination of $u_\tau$, the measurements in FOV2 were subject to additional processing using the sum-of-correlation algorithm with 12 \texttimes\ 3 pixels\textsuperscript{2} Gaussian interrogation windows at 75\% overlap. The resultant velocity fields of the two cameras were combined to obtain the 2D3C velocity field by a least squares fitting approach. The average stereo reconstruction error was approximately 1 pixel. 

From FOV2, wall-normal profiles of the turbulence statistics were obtained by ensemble averaging (denoted by $\langle.\rangle$), in time and streamwise direction, spanning $0.5 \leq x/\lambda_x \leq 1.5$. This method allows us to assess the integral effect of actuation over one streamwise actuation wavelength. The wall-normal profiles are thus averaged over 1000 points in time, and subsequently over 623 points in the streamwise direction, for a total of 623 000 samples per wall-normal point. The uncertainty of the profiles was quantified following the method of \citet{sciacchitano_PIV_2016}. The number of uncorrelated samples, i.e. every fourth element in the streamwise direction to account for the 75\% overlap. The uncertainties for the non-actuated case, at $y^+ = 15$, are 0.069\% and 0.36\% for the mean streamwise velocity and streamwise Reynolds stress, respectively.

\begin{table*}[h]
\caption{Measurement parameters of the SPIV experiments}
\label{tab:params}

\begin{tabularx}{\textwidth}{Xcc}
\hline \hline
Seeding                                       & \multicolumn{2}{c}{1 \textmu m water-glycol droplets}                                                                         \\
Illumination                                  & \multicolumn{2}{c}{Nd:YAG, 532 nm, 2 × 200 mJ (50\% power setting)}                                                                                 \\
Cameras                                       & \multicolumn{2}{c}{sCMOS, 2560 × 2160 pixels\textsuperscript{2}, 6.5 \textmu m pixel, 16 bits}                                                 \\
Camera orientation ($^\circ$)                                      & \multicolumn{2}{c}{$\pm 30$ (with respect to laser-sheet normal)}                                                 \\
Number of recordings                                    & \multicolumn{2}{c}{1000}                                                 \\
Recording frequency (Hz)                      & \multicolumn{2}{c}{10}                                                                                                 \\
Pulse delay (\textmu s)                      & \multicolumn{2}{c}{23}                                                                                                 \\
Aperture                                      & \multicolumn{2}{c}{f/11} \\ 
Post-processing method                        & \multicolumn{2}{c}{Cross-correlation}          \\
Interrogation window size (pixels\textsuperscript{2}) & \multicolumn{2}{c}{16 \texttimes\ 16, circular, 75\% overlap }
\\ \hline
                                              & FOV1                                  & FOV2    \\ \hline
Focal length (mm)                             & 200                                   & 400 (200mm lenses, 2x teleconverters)                 \\
Field of view (mm\textsuperscript{2})                   & 46.5 \texttimes\ 17.5                 & 24 \texttimes\ 14.5                                                       \\
Resolution (pixels/mm)                            & 60                                    & 113                                                        \\
Magnification factor                           & 0.39                                  & 0.73                                                       \\
Observation distance (mm)                     & 510                                   & 540                                                        \\
Vector pitch (mm)                             & 0.067                                 & 0.035                                                        \\
Vector pitch ($\nu/u_{\tau0}$)         & 1.2                                   & 0.64                           \\
Calibration fit error (pixels) & 0.029 & 0.245\\
Stereo reconstruction error (pixels) & 1.00 & 1.00 \\
\hline \hline
\end{tabularx}
\end{table*}
\subsection{Boundary layer fitting and performance metrics}
\label{sec:method:subsec:BLfit_DR}
To obtain the viscous scaling characteristics under drag-reduced conditions, the measured velocity data in the region of $2 \leq y^*\leq 5$ is fitted to the theoretical relation $u^* = y^*$. Here, $u_\tau$ and a wall-normal offset $\Delta y$ are used as fitting variables. To account for the change in $\Delta y$, the bounds of the fit are updated in an iterative manner until convergence; usually 2-3 iterations were found to be sufficient. The maximum standard error of $u_\tau$ was 0.55\%. For the non-actuated case, $u_{\tau0}$ agrees within 0.98\% to the value obtained from using the composite profile of \citet{chauhan_criteria_2009}, using the log-layer parameters of $\kappa = 0.384$ and $B = 4.17$ \citep{nagib_variations_2008}. Further details regarding the fitting accuracy are elucidated in \S~\ref{sec:performance}.
The DR, defined as the reduction in mean wall-shear stress, can now be expressed in terms of $u_\tau$ according to:

\begin{equation}
    \text{DR (\%)} = \frac{\tau_{w0} - \tau_w}{\tau_{w0}}\times 100 = \left(1- \left( \frac{u_\tau}{u_{\tau 0}}\right)^2\right)\times 100.
\end{equation}

The theoretical power requirement for the forcing, given in eq.~\ref{eq:Pin}, is governed by the spanwise wall shear stress required to drive the Stokes layer. The input power is subsequently normalised by the power to drive the non-actuated flow. 
\begin{equation}
    \label{eq:Pin}
    \text{P}_\text{in}\text{ (\%)} = \frac{\int_x \nu \left.(\partial \tilde{w}/\partial y) \right|_{y = 0} A \mathrm{d}x}{\int_x u_{\tau0}^2U_\infty \mathrm{d}x}\times 100.
\end{equation}
The NPS given in eq.~\ref{eq:NPS}, is then defined as the difference between the power saving in terms of DR and the input power. Eq.~\ref{eq:Pin} reveals a quadratic proportionality with amplitude, i.e. $\text{P}_\text{in} \propto A^{+2}$, which explains that the NPS shows an optimum as a function of $A^+$ when assessing the NPS, compared to a monotonically increasing DR trend.

\begin{equation}
    \label{eq:NPS}
    \text{NPS (\%)} = \text{DR} -\text{P}_\text{in}.
\end{equation}

\subsection{Laminar solution of the spatial Stokes layer for an arbitrary waveform}
\label{sec:SSL}
An extension of the laminar SSL model is required to allow for the representation of an arbitrary waveform of the spanwise wall velocity. The classical Stokes layer solution is derived from the $z$-momentum equation of the laminar boundary layer equations, under the assumption that the characteristic wall-normal length scale ($\delta_x$) is small compared to the outer scale. Full details on the derivation can be found in \citet{viotti_streamwise_2009}. The solution corresponding to a harmonic actuation waveform reads as follows:

\begin{equation}
\label{eq:SSL}
\tilde{w}(x,y) = A C_x \mathrm{Re} \left[ e^{\mathrm{i} k_x x} \mathrm{Ai} \left(-\frac{\mathrm{i} y}{\delta_x} e^{-\mathrm{i}4\pi/3}\right)   \right],
\end{equation}
with
\begin{equation}
    \delta_x = \left( \frac{\nu}{k_x u_{y,0}}\right)^{1/3},
\end{equation}
where $C_x = \mathrm{Ai}(0)^{-1}$ is a normalisation constant,  $\mathrm{Ai}$ is the Airy function of the first kind, $\mathrm{Re}$ indicates the real-valued component, $\mathrm{i}$ is the imaginary unit, $k_x$ is the streamwise wavenumber and $u_{y,0}$ is the slope of the streamwise velocity profile at the wall. We now adapt the SSL solution to account for an arbitrary waveform, following the approach taken by \citet{cimarelli_prediction_2013} for the TSL. The linear nature of the governing equation allows for a superposition of various harmonics to make up the desired waveform. In line with their approach, the waveform describing the wall motion is expressed in terms of a Fourier series:
\begin{equation}
    W_w(x) = A\sum_{n=-\infty}^{+\infty} B_n e^{\mathrm{i} k_xn x}, 
\end{equation}
where $B_n$ is the complex coefficient and $k_x n$ the wavenumber, associated with the n\textsuperscript{th} Fourier mode. The corresponding penetration depth is defined as:
\begin{equation}
    \delta_{x,n} = \left( \frac{\nu}{k_x n u_{y,0}}\right)^{1/3} = \delta_xn^{-1/3}.
\end{equation}
Substituting this and superimposing the elementary solutions, the SSL for an arbitrary waveform is then found as follows:
\begin{equation}
\label{eq:SSL_general}
\begin{split}
&\tilde{w}(x,y) = \\
& A C_x \sum_{n=-\infty}^{+\infty} \mathrm{Re} \left[ B_n  e^{ik_x n x} \mathrm{Ai} \left(-\frac{i y}{\delta_x n^{-1/3}} e^{-i4\pi/3}\right)\right].
\end{split}
\end{equation}
\section{Experimental results and discussion}
The SSES was used in a configuration where the belts move in alternating spanwise direction, resulting in the sequence of two square waveforms with $\lambda_x = 22$ mm, and a viscous wavelength of $\lambda_x^+ = 397$. The corresponding actuation parameters are presented in \autoref{tab:actuations}. The belt configuration indicates the spanwise motion direction of each of the four belts, according to the rotation direction of the driving system. The $+$ and $-$ sign dictate a respective positive and negative rotation. The first section discusses the modification of the turbulence statistics resulting from spanwise forcing, examining both $x-y$ scalar fields and wall-normal profiles that result from integration over one streamwise phase. An additional case is considered to highlight the importance of periodicity in the spanwise forcing direction. To address this point, we include a configuration with constant belt movement in the positive spanwise direction, i.e. the configuration $ +\ +\ +\ +$. For the nominal configuration ($+\ -\ +\ -$) we elaborate on the model of the modified SSL and compare it to the experimental results. Finally, we present the performance characteristics of the experiment in terms of DR and theoretical NPS.

\begin{table}[h]
\caption{Overview of the belt configurations and their corresponding actuation parameters.  }
\begin{tabular}{lllll}
\hline \hline
Configuration & $A$ (m/s) & $\lambda_x$ (mm) & $A^+$ & $\lambda_x^+$ \\ \hline
Non-actuated  & -      & -               & -   & - \\
$+\ -\ +\ -$    & 0.57      & 22               & 2.1   & 397           \\
$+\ -\ +\ -$    & 1.13      & 22               & 4.3   & 397           \\
$+\ -\ +\ -$  & 1.70      & 22               & 6.4   & 397           \\
$+\ -\ +\ -$     & 2.27      & 22               & 8.5   & 397           \\
$+\ -\ +\ -$     & 2.83      & 22               & 10.6  & 397           \\
$+\ -\ +\ -$    & 3.40      & 22               & 12.7  & 397           \\
$+\ +\ +\ +$ & 3.40      & -              & 12.7  & -           \\
\hline \hline
\end{tabular}

\label{tab:actuations}
\end{table}
\subsection{Turbulence statistics}
We discuss the actuation impact on the turbulence statistics at the hand of streamwise-wall-normal scalar fields in FOV1, comparing the non-actuated case with the actuated case at $A^+ = 12.7$. Subsequently, the ensemble-averaged wall-normal 
profiles over the central phase from FOV2 are presented. Under periodic forcing, the total spanwise stress is decomposed into a stochastic component ($w''$) and a phase-wise component ($\tilde{w}$), according to $ w' = w'' + \tilde{w}$. Here the phase-wise component corresponds to the spanwise velocity component induced by the spanwise forcing (i.e. the SSL). The turbulence statistics profiles also reveal the trends with spanwise velocity amplitude, $A^+ \approx 2 - 12$. To highlight the absolute changes with respect to the non-actuated case, reference scaling with $u_{\tau0}$ is applied unless otherwise specified.

\begin{figure*}
    \includegraphics[]{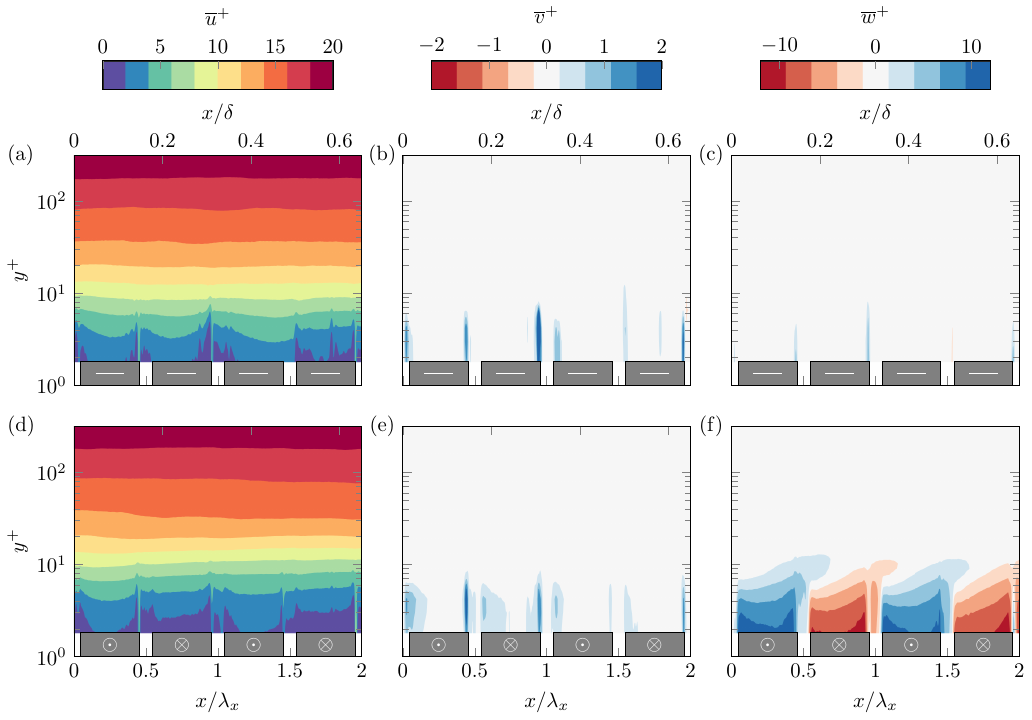}
    \caption{Scalar fields of the mean streamwise velocity (a,d), the mean wall-normal velocity (b,e) and the mean spanwise velocity (c,f), for the non-actuated case (top row) and actuated case (bottom row) at $A^+ = 12.7$. The grey shaded regions indicate the belt locations and their respective in-plane moving direction; $-$ indicates no rotation, whereas $\odot$ and $\otimes$ denote the positive and negative spanwise direction, respectively. }
    \label{fig:ScalarfieldsMean}
\end{figure*}

\begin{figure*}
    \includegraphics[]{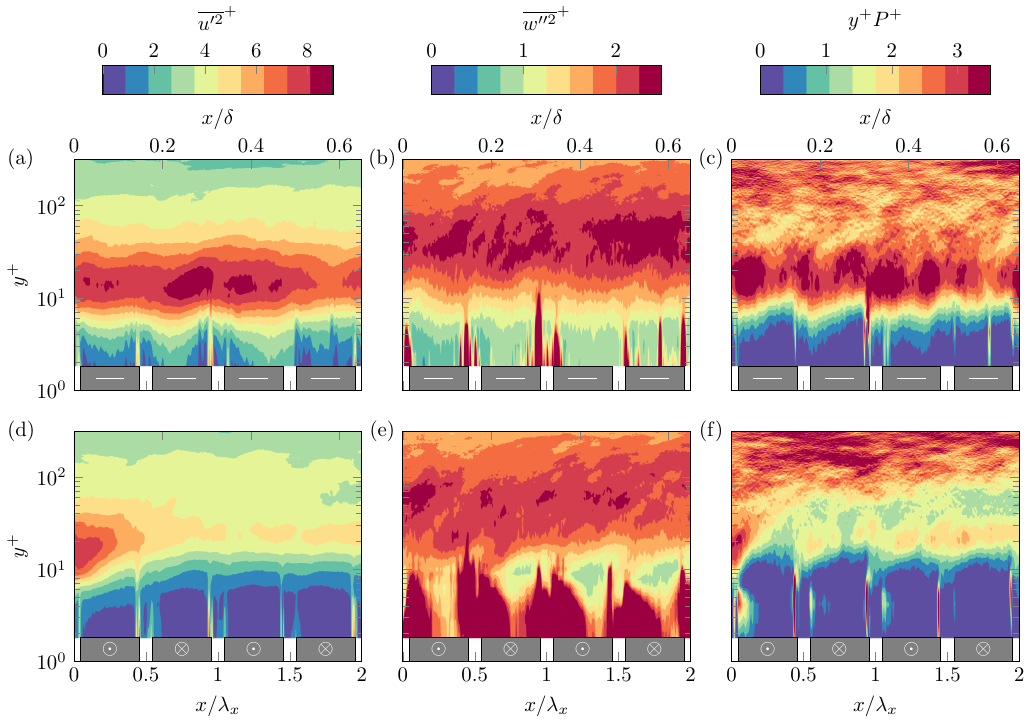}
    \caption{Scalar fields of the streamwise stress (a,d), the stochastic spanwise stress (b,e) and the pre-multiplied production of turbulence kinetic energy (c,f), for the non-actuated case (top row) and actuated case (bottom row) at $A^+ = 12.7$.}
    \label{fig:ScalarfieldsHO}
\end{figure*}

\begin{figure*}
    \includegraphics[]{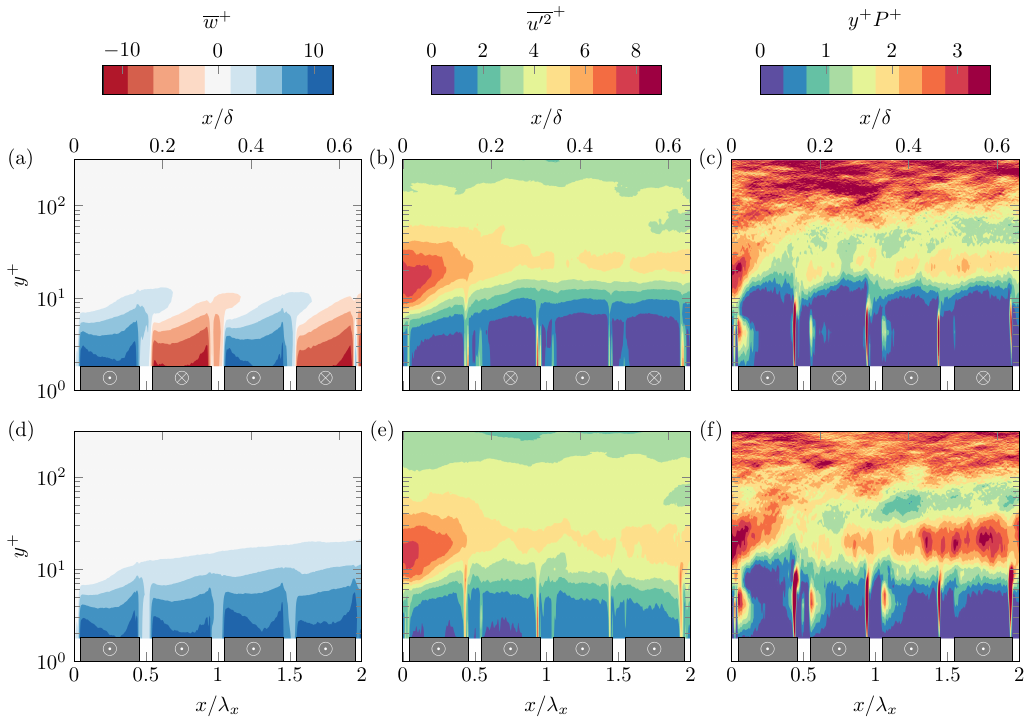}
    \caption{Scalar fields of the mean spanwise velocity (a,d), the streamwise stress (b,e) and the pre-multiplied production of turbulence kinetic energy (c,f), for the nominal configurations: $+\ -\ +\ -$ (top row) and constant forcing $+\ +\ +\ +$ (bottom row) at $A^+ = 12.7$.}
    \label{fig:ScalarfieldsAltWf}
\end{figure*}

\subsubsection{Streamwise-wall-normal scalar fields}
The scalar fields of the turbulence statistics, in terms of the three mean velocity components and a selection of higher-order turbulence statistics, are presented in Fig.~\ref{fig:ScalarfieldsMean} and Fig.~\ref{fig:ScalarfieldsHO}. A general remark applies to all scalar fields, which highlight the presence of spurious measurements in the near-wall region, reflected by the strong peaks centred around the belts' leading and trailing edges. It is postulated that these peaks arise from unsteady laser-light reflections on the edges of the spanwise slots of the aluminium surface plate through which the belts traverse. Despite attempts to remove these artefacts, the reflections could not be filtered completely from the SPIV images in image pre-processing, consequently impacting measurement results in the near vicinity of the wall. The spanwise stress is particularly affected, which is believed to originate from a slight variation in the spanwise location of the reflection between image pairs, while maintaining an approximate fixed $x-y$ position, thus primarily correlating as spurious spanwise fluctuations. Nevertheless, the spurious data is contained very close to the wall, in the region $y^+ \lessapprox 10$, while the other scalar fields are minimally affected. This is also revealed from the turbulence statistics profiles in Fig.~\ref{fig:turbulence statistics}: when streamwise averaging is applied, only the spanwise stresses in the region $y^+ \leq 10$ are significantly increased. For completeness and transparency about the experimental data, we include all data $y^+\geq 2$. However, we stress that the data in the affected regions should be regarded with caution, especially the spanwise stress. 

The scalar fields of the mean streamwise velocity are depicted in Fig.~\ref{fig:ScalarfieldsMean} (a,d), for the non-actuated and actuated case, respectively. In the near-wall region, $y^+ \leq 15$, the non-actuated case exhibits a heterogeneous distribution of the streamwise velocity, reflecting minor changes in local skin friction. This variation is believed to result from the experimental implementation, involving step changes in roughness (i.e. the belts) and the $0.9\nu/u_{\tau0}$ wide spanwise grooves, which can slightly modify the local skin friction. Under actuation, the near-wall contour lines show an upward movement in the streamwise direction, indicating a ``thickening" of the viscous sublayer and buffer layer. This observation aligns with earlier findings on spanwise forcing \citep{baron_turbulent_1996}. 
Fig.~\ref{fig:ScalarfieldsMean} (b,e) displays the mean wall-normal velocity. Apart from the previously mentioned spurious measurements, a nominally zero mean flow is observed. Similarly, a zero mean spanwise velocity can be observed, see Fig.~\ref{fig:ScalarfieldsMean} (c,f), except for the near-wall flow in the actuated case. Here, positive and negative contours reflect the spanwise shear layer resulting from the alternating spanwise wall movement. Centred over the belts, the contours are proportional to $\pm A^+$, with the spanwise velocity decreasing when moving in the positive wall-normal direction, reaching approximately zero at $y^+ \approx 25$. Additionally, the contours show a forward inclination. This behaviour qualitatively aligns with scalar fields of the Stokes layer, as discussed by \citet{viotti_streamwise_2009} and \citet{Agostini_spanwise_2014}. Further discussion on the spanwise velocity profiles and a comparison to the modified SSL model is presented in \S~\ref{sec:SSL}.

Fig.~\ref{fig:ScalarfieldsHO} (a,d) shows the scalar fields for the streamwise Reynolds stress. In the non-actuated case, a distinct energetic peak is evident at the characteristic location $y^+ \approx 15$. Despite some minor variation, the energy peak remains constant in magnitude, location, and peak width, over the streamwise extent. Upon actuation, the stress is minimally affected upstream of the actuation surface. In the streamwise direction, however, a significant reduction in the near-wall stress peak is observed over the first belt already. This reduction is followed by a more gradual transition over the second belt to an equilibrium energy state from $x/\lambda_x \approx 0.75$ onward. Accordingly, the streamwise stress peak is shifted in the positive wall-normal direction, which is consistent with earlier observations by \citet{ricco_effects_2004}. Examining the (stochastic component of the) spanwise stress in Fig.~\ref{fig:ScalarfieldsHO} (b,e), a moderate reduction in energy is noted in the logarithmic region, with the energetic peak shifting to a higher wall-normal location. 

Lastly, we consider the production of turbulence kinetic energy (TKE), which is defined as:
\begin{equation}
\label{eq:prod}
    P_{ij} = -\overline{u_i'u_j'} \frac{\partial\overline{u_i}}{\partial x_j} \approx  -\overline{u'v'}\frac{\partial \overline{u}}{\partial y}.
\end{equation}
In a non-actuated boundary layer, assuming spanwise homogeneity and quasi-homogeneous development in the streamwise direction (i.e. $\partial/\partial z = 0$, $\partial/\partial x\approx 0$), coupled with the approximations $\overline{v} = \overline{w} = 0$, the TKE production simplifies to $P = -\overline{u'v'}\partial \overline{u}/\partial y$. When spanwise forcing is introduced ($\partial/\partial x\neq 0$ and $\overline{w} \neq 0$), the additional production components for the actuated case were found to be an order of magnitude lower in regions unaffected by spurious fluctuations. Consequently, we only consider the dominant production term expressed by Eq.~\ref{eq:prod}, for both non-actuated and actuated cases. The scalar fields in Fig.~\ref{fig:ScalarfieldsHO} (c,f) depict the premultiplied form of TKE production ($y^+ P^+$). In the non-actuated case, the production exhibits the characteristic peak in the near-wall region. Under actuation, the response is similar to that of streamwise stress $\overline{u'^2}$, albeit with an even shorter spatial transient. The production is significantly reduced over the first belt, and the characteristic peak is strongly attenuated and shifted in the positive wall-normal direction. An approximately steady-state response is established from the second belt onward.

\subsubsection{Modification of the turbulence statistics by constant spanwise forcing}
We have discussed the modification of the first- and second-order turbulence statistics by steady spatial periodic forcing. To highlight the importance of periodicity in spanwise forcing direction, the following subsection will elucidate the qualitative response of the turbulence statistics to constant spanwise forcing, i.e. when all belts are moving in the same spanwise direction. Furthermore, this configuration is closest to the experiment of \citet{kiesow_near-wall_2003}, who investigated the physics associated wait a planar shear-driven three-dimensional turbulent boundary layer by employing a single spanwise running belt. Contrary to the findings for spatial periodic forcing, the authors observe an increase in both normal and Reynolds shear stresses, associated to an increase in TKE production. It should however be noted that there are distinct differences to their work: firstly, their larger 13 cm streamwise extent of forcing, and actuation at higher (outer-scaled) spanwise velocities of $W_w/U_\infty = 1-2.75$. Furthermore, the turbulence statistics were only investigated in the region surrounding the trailing edge of the actuation surface, whereas our work is focused on the initial transient. 

Fig.~\ref{fig:ScalarfieldsAltWf} depicts the scalar fields of the mean spanwise velocity, the streamwise normal stress and the premultiplied TKE production, for the previously discussed case of alternating forcing and the case with constant spanwise forcing. The contours of spanwise velocity Fig.~\ref{fig:ScalarfieldsAltWf} (d) show the effect of the non-perfect waveform when a multitude of belts is rotated in the same direction. In the 2 mm transition region between belts, close to the wall, the spanwise velocity partially recovers to a lower value. This effect is not observed above $y^+ \geq 10$, where the contours of the different belts merge. It can also be observed that the spanwise shear layer penetrates to a higher wall-normal location under constant forcing. Assessing the higher-order turbulence statistics (Fig.~\ref{fig:ScalarfieldsAltWf} (e,f)), the initial transient reveals a close match with the initial transient of the spatial periodic case. However, the attenuation is followed by a gradual increase in the streamwise TKE and the production of TKE. This is in contrast to the steady energy state obtained for the oscillatory forcing. The difference can be explained by the absence of a significantly strong Stokes strain rate, following the notion of \citet{Bradshaw_measuements_1985} and \citet{Agostini_spanwise_2014}. It is hypothesised that over a long enough streamwise extent the turbulence statistics will reestablish to their non-actuated state, and possibly increase, as in the experiment by \citet{kiesow_near-wall_2003}. However this fact cannot be verified, due to the limited spatial extent of the actuation surface and the SPIV experiment, and is deemed outside of the scope of the current investigation.

\begin{figure*}[t]
\includegraphics[]{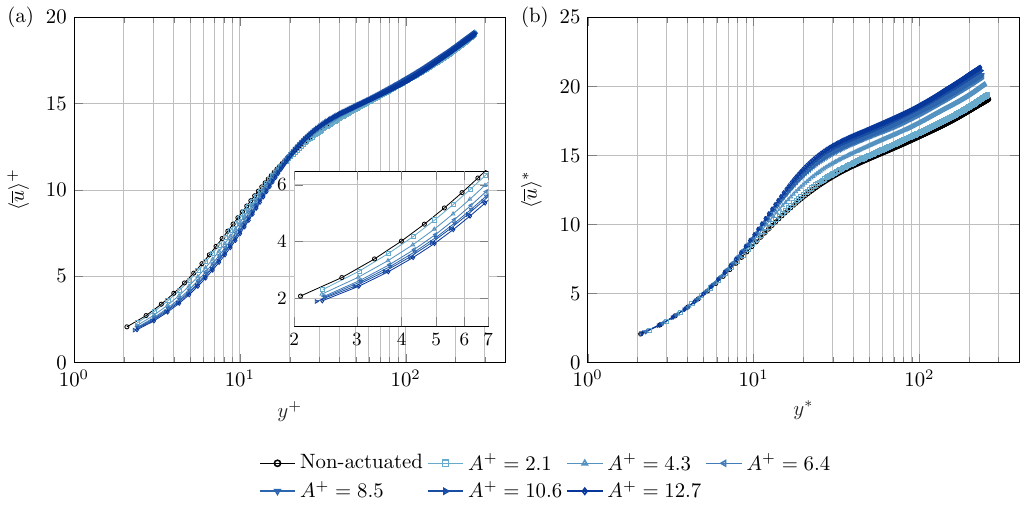}
    \caption{Mean streamwise velocity profiles (a) scaled with $u_{\tau0}$, (b) scaled with $u_{\tau}$.}
    \label{fig:u_mean}
\end{figure*}

\begin{figure*}[t!]
    \includegraphics[]{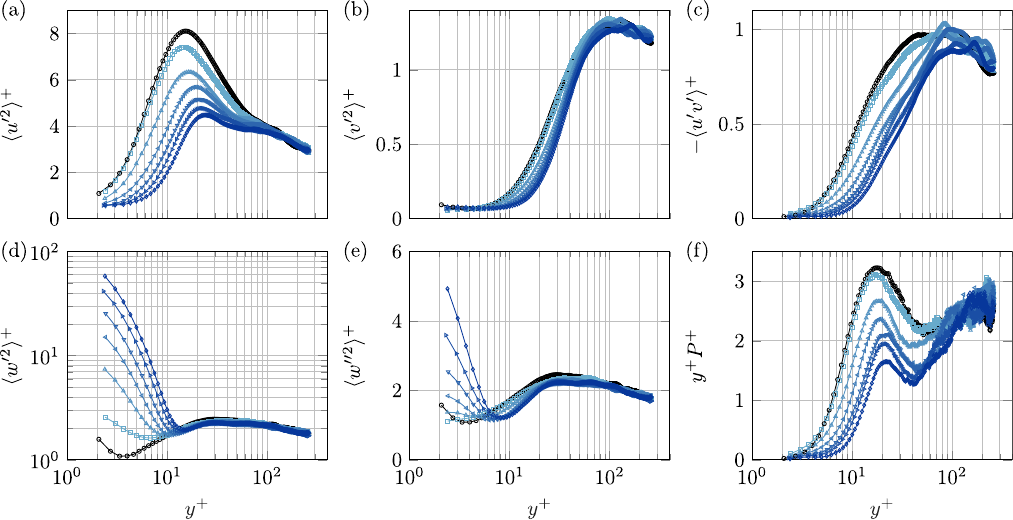}
    \caption{(a,b) Streamwise and wall-normal Reynolds stresses and (c)  Reynolds shear stress. Spanwise stress is decomposed in (d) total spanwise stress and (e) the stochastic component. (f) Premultiplied production of turbulence kinetic energy. Legend is depicted in Fig.~\ref{fig:u_mean}.}
    \label{fig:turbulence statistics}
\end{figure*}
\subsubsection{Wall-normal profiles}
To further quantify the flow modulation effect of the spanwise forcing on the near-wall turbulence, we present in this section wall-normal profiles of various turbulence statistics. We consider the case $\lambda_x^+ = 397$, with increasing $A^+ = 2.1-12.7$. The statistics are obtained from FOV2, zoomed in on the central two belts, by streamwise averaging over one complete phase, i.e. $0.5 \leq x/\lambda_x\leq 1.5$; the latter is indicated by the $\langle . \rangle$ operator. From the previously discussed scalar fields, it can be concluded that the turbulence statistics show a minimal initial transient in the considered streamwise domain. Consequently, this methodology allows the assessment of the integral impact of actuation over one streamwise actuation wavelength.

The mean streamwise velocity profiles are shown in Fig.~\ref{fig:u_mean} (a,b), in scaling with either the reference or the actual friction velocity, respectively. Scaling with $u_{\tau0}$ reveals a reduction of mean velocity below $y^+ \leq 15$, which is positively correlated with $A^+$. This reflects a reduction of the near-wall velocity gradient, directly indicative of a drag-reduced state. The behaviour is consistent with previous studies on spanwise forcing \citep{ricco_effects_2004, gatti_reynolds-number_2016}. To further emphasise this effect, scaling with $u_\tau$ is considered. The profiles now show similarity in the viscous sublayer, with an upward shift (i.e. $\Delta B > 0$) of the logarithmic region under actuation, which is again consistent with a reduction in skin friction, as will be discussed in \S~\ref{sec:performance}. The velocity profiles, specifically noticeable in Fig.~\ref{fig:u_mean} (a), reveal an overshoot of the velocity profile in the region $20 \leq y^+ \leq 40$ compared to the non-actuated case. These results are in line with the conceptual model of \citet{choi_near-wall_2002}, which suggests that this modification of the streamwise velocity profile is due to an induced negative spanwise vorticity. This behaviour is also reflected by the converging contour lines in Fig.~\ref{fig:ScalarfieldsMean} (d) for the range $10\leq \overline{u}^+\leq 15$.

The higher-order turbulence statistics in terms of the variance profiles for the three velocity components, the Reynolds shear stress, and the TKE production are depicted in Fig.~\ref{fig:turbulence statistics}. Assessing Fig.~\ref{fig:turbulence statistics} (a), we observe that upon actuation the streamwise stress is attenuated in the region up to $y^+ \approx 100$, and that the near-wall peak is shifted to higher wall-normal positions. In line with the DR behaviour inferred from the mean velocity profiles (\S~\ref{sec:performance}), the effect is correlated with $A^+$. The most significant reduction is found in the near-wall stress peak, which is centred around $y^+ = 15$ for the non-actuated case. The near-wall peak shows a 45\% reduction in magnitude for the maximum value of $A^+$ considered. This response is consistent with earlier experimental observations of drag-reduced flow under spanwise forcing \citep{laadhari_turbulence_1994, ricco_effects_2004}. The stresses $\langle v'^2 \rangle^+$ and $-\langle u'v' \rangle^+$, in Fig.~\ref{fig:turbulence statistics} (b,c), reflect similar behaviour, where the profiles shift to higher wall-normal locations and attenuation is most significant close to the wall.

The total spanwise stress, in Fig.~\ref{fig:turbulence statistics} (d), shows a strong increase in the near-wall region (note the log-log scale), as a direct consequence of the phase-wise fluctuating component $\langle \tilde{w}^2 \rangle^+$. This component corresponds to the streamwise variation of the Stokes profiles, hence, the peak's magnitude is proportional to $A^{+2}$. When the phase-wise component is subtracted, the stochastic stress is obtained, which is depicted in Fig.~\ref{fig:turbulence statistics} (e). This component also highlights the influence of the aforementioned spurious fluctuations, which strongly increase the apparent fluctuations in the region $y^+ \leq 10$ and which is correlated to the actuation amplitude. In agreement with the other stresses, we observe that under actuation, the $\langle w''^2 \rangle^+$ profiles are shifted in the wall-normal direction and show a minor energy reduction up to a height of about $y^+ \approx 100$. 

Lastly, we discuss the premultiplied TKE production, in Fig.~\ref{fig:turbulence statistics} (f). Actuation significantly reduces the peak in the buffer layer, and again, the profiles are shifted to a higher wall-normal position, similar to the streamwise Reynolds stress. For $A^+ = 12.6$, a reduction of integral TKE production, over the considered wall-normal region, of 39\% is found.

\subsection{Comparison to the modified spatial Stokes layer}
We now address how the experimentally realised forcing relates to the theoretical Stokes layer by comparing measured spanwise profiles to the model of the modified SSL derived in \S~\ref{sec:SSL}. The periodic waveform, consisting of two 9 mm regions actuated at $\pm A$, is prescribed by ten single-sided Fourier coefficients. The 2 mm transition between belts is approximated by a linear increase from $-A$ to $A$ to prevent Gibbs phenomena. We investigate the case for $A^+ = 12.7$ from FOV2, discuss the scalar field of $\tilde{w}$, and subsequently the spanwise velocity profiles at various locations. The half-phase ($\widehat{\lambda}_x = \lambda_x/2$) is used to normalise the streamwise location, to ensure that the same value after the decimal point indicates the same location on the half-phase.

\begin{figure}
    \includegraphics[]{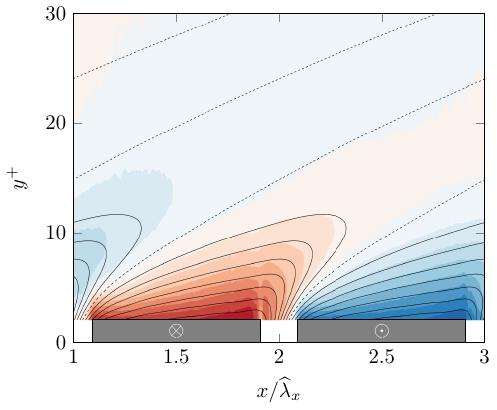}
    \caption{Contours of $\tilde{w}/A$ at $A^+ = 12.7$, for the experiments (filled) and the theoretical model of the modified SSL (black lines), contour levels range $\pm1$ with a level step of 0.1, zero-contours are indicated by the dashed line.}
    \label{fig:SSL_SF}
\end{figure}

\begin{figure*}
\includegraphics[]{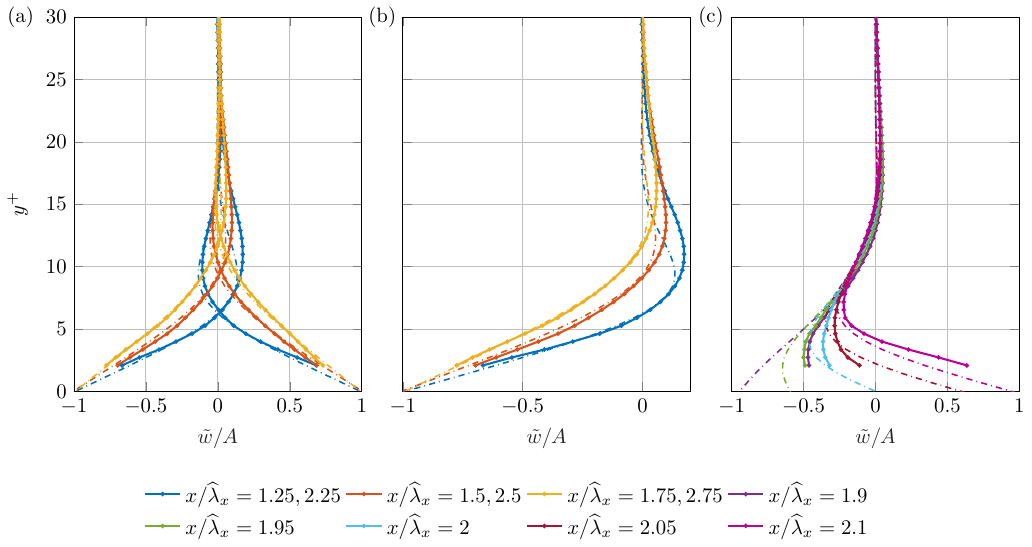}
\caption{Spanwise velocity profiles at $A^+ = 12.7$, where the colour denotes the location on the half-phase. Experimental data plotted with solid lines, modified SSL displayed for the same colour indicated with a dashed-dotted line. (a) Profiles over the belts at three locations per half-phase, (b) a detailed view of the profiles over the first half-phase, (c) five linearly spaced profiles in the transition region between belts.}
\label{fig:SSL}
\end{figure*}

Fig.~\ref{fig:SSL_SF} shows contours of $\tilde{w}/A$ as obtained from the experiments, while the black contour lines depict the analytical model of the modified SSL, for one phase. The figure reveals a strong qualitative agreement between the model and the experimental data. Close to the wall ($y^+ \leq 10$), the contours $\pm \tilde{w}/A \geq 0.3$ exhibit good agreement between the experimental results and the theoretical model. However, in the region $y^+ \geq 10$, the experiment diverges from the theoretical model. Specifically, the contour $\pm 0.1$ pertains to higher wall-normal heights and has an increased streamwise inclination.

To further quantify the response of the Stokes layer, spanwise velocity profiles extracted at three locations for each half-phase (i.e. belt) are presented in Fig.~\ref{fig:SSL} (a). The experimental profiles are subjected to streamwise averaging across a $x \pm 0.1$ mm range. It is observed that the profiles resemble a clear Stokes-like layer, with a steep spanwise velocity gradient at the wall, followed by an overshoot around $y^+ \approx 10$, after which it decays to zero at $y^+ \approx 25$. This behaviour qualitatively aligns with the SSL model, subject to equation~\ref{eq:SW}, at moderate wavelengths \citep{quadrio_laminar_2011, viotti_streamwise_2009}. Some small degree of asymmetry between the two half-phases is noticeable, possibly attributed to a spatial transient of the SSL, which has only experienced actuation over one half-phase upstream of the FOV. This behaviour can also be observed from the scalar field in Fig.~\ref{fig:SSL_SF}.

To make a comparison to the theoretical model for the modified SSL, the first half-phase is highlighted in more detail in Fig~\ref{fig:SSL} (b). The experimental profiles match the theoretical model quite well, with an almost exact match in the region $y^+ \leq 10$. The slope of the spanwise velocity profile is in line with the model and decreases when moving downstream over the belt. Hence, the model can be used to estimate the power input required to drive the spanwise flow, which is calculated from the integral spanwise wall shear stress (i.e. $\propto \left.(\mathrm{d}\tilde{w}/\mathrm{d}y)\right|_{y=0}$). We used this approach to estimate the input power, allowing the calculation of the NPS; a further discussion on these outcomes is presented in \S~\ref{sec:performance}. As was shown in the scalar field of Fig.~\ref{fig:SSL_SF} the experimental data beyond $y^+ = 10$ start to divert from the theory. The overshoot of the experimental profile is larger in magnitude, present at higher wall-normal locations, and penetrates deeper into the boundary layer. These differences may be attributed to the effect of turbulent flow further away from the wall, in contrast to the laminar flow assumption of the SSL model. 

The idealised model furthermore deviates from the practical realisation by linearly transitioning between half-phases, instead of the abrupt changes and the zero spanwise wall velocity in the transition region. To assess the effect of this choice, five linearly spaced profiles spanning $1.9 \leq x/\widehat{\lambda}_x \leq 2.1$ in the transition region are studied, as depicted in Fig~\ref{fig:SSL} (c). In the region $y^+ \leq 10$ a good match between the model and the experiment can be observed. Close to the wall some variation between the two can be observed, but it should be noted that the transition region below $y^+ \leq 10$ is the region affected by the spurious measurements. The penetration depth is defined as the wall-normal location where the phase-wise standard deviation of $\tilde{w}$ reduces to $Ae^{-1}$. The model and experimental values are $\delta_{s, mod}^+ = 5.0$ and $\delta_{s, exp}^+ = 4.7$, respectively. At first glance this result appears conflicting, since the experimental profile, specifically the overshoot, penetrates deeper into the boundary layer. However, following the above definition, the penetration depth is confined to the viscous sublayer. In this region the model shows a slightly higher variance, explaining the small difference. Since there are only minor differences, we conclude that the current way of modelling the Stokes layer can be applied with reasonable confidence to estimate the power requirement to drive the spanwise flow.

\subsection{Performance characteristics}
\label{sec:performance}
\begin{figure*}
    \includegraphics[]{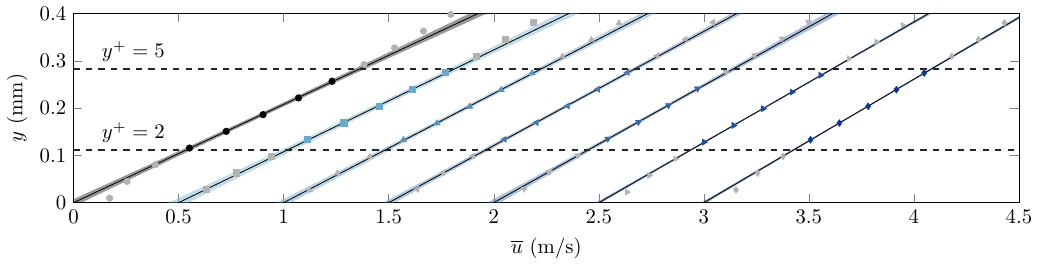}
    \caption{Near-wall streamwise velocity data and their corresponding linear fit. From left to right are the non-actuated case and the actuated cases $A^+ = 2.1-12.7$ in increasing order, the previous  colour scheme and markers are adopted (legend is depicted in Fig.~\ref{fig:u_mean}), and the cases are spaced with an increment of $\Delta\overline{u} = 0.5$ m/s. Dashed horizontal lines indicate the fitting domain $2 \leq y^+\leq 5$. The linear fit is indicated by the solid black lines, the shaded area represents the 95\% confidence bounds.}
    \label{fig:DR_Fit}
\end{figure*}
\begin{figure}
\includegraphics[]{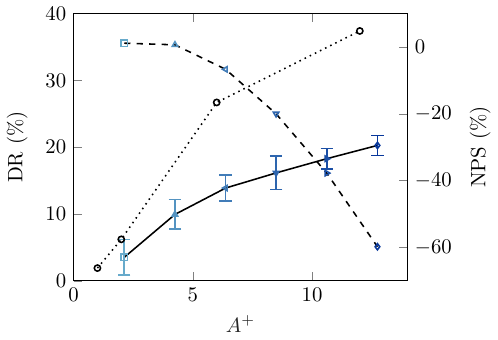}
    \caption{DR (solid) and NPS (dashed) vs $A^+$ compared to the DR of \citet{viotti_streamwise_2009} for $\lambda_x^+ = 312$ (dotted), error bars indicate the 95\% confidence interval of DR.}
    \label{fig:performanance}
\end{figure}

Lastly, we discuss the drag-reducing performance achieved with the experimental setup, we consider the cases associated with $\lambda_x^+ = 397$. Starting with the accuracy of the linear fit used to obtain $u_\tau$, which defines the DR. In Fig~\ref{fig:DR_Fit}, we present the near-wall velocity data, which is employed for a linear fit within the $2 \leq y^* \leq 5$ region, following the methodology outlined in Section~\ref{sec:method:subsec:BLfit_DR}. The fitting domain, containing 4-5 data points, was selected to capture a significant portion of the linear region while excluding wall-biased data. The theoretical linear relation $u^* = y^*$ starts to deviate from the mean velocity from $y^+ \approx 4$ onwards. The selection of the domain $2 \leq y^* \leq 5$ was made as a trade-off between accuracy and uncertainty of the fit (i.e. number of points). In addition, we observed no significant differences in the fit (hence, DR) when employing the smaller domain $2 \leq y^* \leq 4$. It is important to note that for determining the friction velocity, we utilise the streamwise ensemble-averaged velocity profiles. Similar to the approach used for turbulence statistics, this provides us with an integral measure of the DR over one streamwise phase. While the current methodology may not offer the precision required for determining the ``absolute" wall-shear stress, we believe it is suitable to investigate DR in terms of relative differences with respect to the non-actuated case. In future work, more direct methods for determining the wall-shear stress may be considered, such as oil-film interferometry \citep{Tanner_study_1976} or balance measurements \citep{baars_wall-drag_2016}.

The tight 95\% confidence bounds reflect the accuracy of the fit, which improves slightly as $A^+$ increases. The standard error (SE) of the fit for $u_\tau$ ranges from 0.55\% to 0.23\% from left to right in the figure (i.e. increasing $A^+$). Applying uncertainty propagation, we calculate the SE on the DR as follows:
\begin{equation}
    \label{eq:SE_DR}
    \mathrm{SE}_{\mathrm{DR}} = \sqrt{\left(\frac{-2u_\tau}{u_{\tau0}^2}100\right)^2\mathrm{SE}_{u_\tau}^2 + \left(\frac{-2u_\tau^2}{u_{\tau0}^3}100\right)^2 \mathrm{SE}_{u_{\tau0}}^2}.
\end{equation}
The 95\% confidence interval is calculated using Student's t-distribution for small sample sizes. The sample size is dictated by the degrees of freedom for the error, which is the number of samples minus the order of the fit which is two (i.e. fitting variables $u_\tau$ and $\Delta y$). 

The performance characteristics of the experimental apparatus are quantified in Fig.~\ref{fig:performanance}, where for reference the data of DR from the DNS of \citet{viotti_streamwise_2009} is added, subject to equation~\ref{eq:SW} with $\lambda_x^+ = 312$ at $Re_\tau = 200$. The experimentally obtained DR increases monotonically at a decreasing rate with $A^+$, to a maximum value of 20\%. Qualitatively this behaviour compares well to the DNS reference. The alignment of trends with the DNS reference is a promising validation, offering support to the hypothesis that we are effectively reducing drag through the spanwise forcing mechanism. In absolute terms, the DR is lower than the DNS reference. This is primarily attributed to the expected streamwise transient of DR from the onset of actuation. As shown by \citet{ricco_effects_2004}, the DR rises to a maximum over a streamwise extent of $3-4\delta$, whereas our measurement domain extends over only $0.16\leq x/\delta \leq 0.50$. 
To support this claim, it is conceivable to quantify the streamwise transient of DR. However, this task is considered beyond the scope of the current study, especially in view of the limited streamwise extent of the current setup, and is suggested as an area of exploration in subsequent research. Another factor which negatively impacts the performance, albeit only a minor effect, is the higher Reynolds number for the experimental conditions compared to $Re_\tau = 200$ \citep{gatti_reynolds-number_2016}. Lastly, the experimental realisation of the SW boundary condition may not perform to the same degree as the idealised conditions imposed in the numerical studies.

The NPS is also indicated on the right axis of Fig.~\ref{fig:performanance}. It should be noted that this is the theoretical NPS to drive the Stokes layer, hence it does not take into account any mechanical losses of the driving system. As a function of $A^+$, the NPS reveals a decreasing trend, at an increasing rate. A positive NPS is only observed at the lowest two amplitudes, with a maximum of 1\% at $A^+ = 2.1$. The maximum is followed by a sharp decline to large negative values, reaching a minimum of -60\% for $A^+ = 12.7$. The behaviour results from the balance between DR and input power ($\text{P}_\text{in} \propto A^{+2}$). 
The NPS is essentially zero or negative, whereas \citet{viotti_streamwise_2009} achieved an optimum of 23\% at $A^+ = 6$. We believe the subpar performance is attributed to an overall lower DR achieved, combined with the higher power requirements associated with square wave forcing \citep{cimarelli_prediction_2013}.

\section{Concluding remarks}
A proof of concept was presented for an experimental realisation of spatial spanwise forcing in an external turbulent boundary layer flow. The development of the experimental apparatus was instigated by the limited attention given to SW forcing so far, particularly in an experimental context. Our concept employs a series of belts, rotating in alternating spanwise direction, to create a spatial square-wave forcing. We presented evidence that a strong flow control effect can be achieved with the current experimental setup, in terms of turbulence statistics and its associated drag-reduction performance metrics. Furthermore, the analytical model of the modified SSL shows good agreement with the experimental square-wave forcing.

Scalar field representations reveal that the turbulence statistics exhibit a short spatial transient: the stresses and production are already strongly attenuated over the first belt, followed by a more gradual decrease to a steady-state energy response over the second belt. This corresponds to a transient extent of the order of $x/\delta \approx 0.25$ for the turbulence statistics, which is much shorter than the transient distance reported in literature for the DR itself (which is of the order of  $3-4\delta$). Under actuation, the streamwise velocity profiles show a reduction in the near-wall region, positively correlated with $A^+$. This indicates a thickening of the viscous sublayer, a qualitative feature of drag-reduced flow, aligning with earlier observations on spanwise forcing. The Reynolds stress profiles reflect attenuation up to a height of $y^+ \approx 100$ and are shifted to higher wall-normal locations. Specifically, the streamwise stress peak shows a reduction of 44\% at the maximum spanwise velocity amplitude. Furthermore, for this case, the integral TKE production is reduced by 39\%.

General agreement between the analytical model of the modified SSL and the experimental profiles was found, showing an almost exact match in the region $y^+ \leq 10$. Furthermore, the penetration depths of the two are in the same order at $\delta_{s, mod}^+ = 5.0$ and $\delta_{s, exp}^+ = 4.7$. This leads us to conclude that the model can be used to estimate the theoretical power input associated with the spanwise forcing. 

Qualitatively, the DR trends are in line with the expected trends from the literature, with a maximum of 20\% within the parameter space investigated. This result gives us confidence that the experimental setup is effectively working on the spanwise forcing mechanism. The lower DR margin is attributed to the limited spatial extent of the current experimental setup. As a result of this lower DR, combined with the elevated power requirements for square-wave forcing, no significant positive NPS is found, with a decline to negative values as $A^+$ is increased.

These findings offer perspective for future research into the working mechanisms responsible for the DR by spanwise forcing. A fundamental understanding of the underlying physics can advance energy efficiency in practical applications, e.g. through its application in passive form \citep{ricco_review_2021}. Extending the streamwise extent of the actuation surface is foreseen as a valuable next step. In such a setup the spatial transient of DR and its full DR potential can be quantified. Wider belts are also advised to realise a larger viscous wavelength of the order of the optimum value. The research may also benefit from a more direct determination of the wall-shear stress, e.g. by oil-film interferometry, to give more confidence in the performance characteristics. Furthermore, the modified SSL theory may hold potential for a predictive model for DR and NPS, similar to those of \citet{choi_drag_2002} and \citet{cimarelli_prediction_2013}. 

\backmatter





\bmhead{Acknowledgements}
This work was financially supported by the Netherlands Enterprise Agency under grant number TSH21002. The authors wish to give special thanks to BerkelaarMRT B.V. for the mechanical design and realisation of the experimental apparatus.

\section*{Declarations}

\bmhead{Conflict of interest}
The authors declare no competing interests.

\bmhead{Authors’ contributions}
Conceptualisation: MK, FH, FS, BvO; Methodology: MK; Formal Analysis: MK; Investigation: MK; Writing – original draft: MK; Writing – review \& editing: All authors; Visualisation: MK; Supervision: FH, FS, BvO; Funding acquisition: OvC, MvN;

\bmhead{Data availability statement} The data is available upon reasonable request to the corresponding author.

\bibliography{bibliography}

\end{document}